\documentclass[
 reprint,
superscriptaddress,
 amsmath,amssymb,
 aps,
prl
]{revtex4-2}

\usepackage{graphicx}
\usepackage{dcolumn}
\usepackage{bm}
\usepackage[version=4]{mhchem}

\begin{document}

\preprint{APS/123-QED}

\title{Towards rational design of power-law rheology via DNA nanostar networks}

\author{Nathaniel Conrad}
\affiliation{Physics Department, University of California, Santa Barbara, CA 93106}
 \altaffiliation[Currently at ]{McKetta Department of Chemical Engineering, The University of Texas at Austin, Austin, TX 78712}

\author{Omar A. Saleh}
\affiliation{Physics Department, University of California, Santa Barbara, CA 93106}
\affiliation{Materials Department, University of California, Santa Barbara, CA 93106}
\affiliation{Biomolecular Science and Engineering Program, University of California, Santa Barbara, CA 93106}

\author{Deborah K. Fygenson}
\affiliation{Physics Department, University of California, Santa Barbara, CA 93106}
\affiliation{Biomolecular Science and Engineering Program, University of California, Santa Barbara, CA 93106}

\email[Corresponding authors' e-mail:]{nconrad@ucsb.edu, saleh@ucsb.edu, fygenson@ucsb.edu}

\date{\today}

\begin{abstract}
We measure the rheology of transient hydrogels comprised of a single type of DNA nanostar that makes both strong and weak bonds. 
These gels exhibit power-law frequency-dependence of their storage and loss moduli, with scaling exponents that depend on the proportions of the two bonds. 
A diffusive stress-relaxation model, in which the strong-bond sub-network relieves stress by diffusing through an effective viscosity imposed by the weak bonds, explains the scaling of their moduli. 
The model has implications for the fractal dimensions of the strong-bond sub-network that are in good agreement with measurements and makes testable predictions for the viscoelasticity of other transient hydrogels.
Overall, this work demonstrates the power of DNA nanotechnology to decipher, and potentially rationally design, power-law rheology.
\end{abstract}

\maketitle



Hydrogels of transiently-bonding molecules/particles are used in applications ranging from industrial (agriculture, cosmetics, printing) to medical (biosensing, drug-delivery, wound healing) where their combination of high water content and viscoelasticity is uniquely enabling \cite{HOARE20081993,Fanalista2019,FIJAN20098,YOSHIMURA199871,Rosales2020,Bush2021,Lattuada2020}.
Understanding how the viscoelasticity of a transient hydrogel derives from its underlying structure and molecular interactions would greatly advance rational engineering and may yield insight into the design and control of complex biomaterials. 

Many transient hydrogels are ``Maxwell materials'' \cite{Rubinstein, Grindy2015,Conrad2019,Katashima2022,Yesilyurt2017, Rosales2020}, meaning they have a single mode of stress relaxation ({\em i.e.}, the breaking of their transient bond), with a characteristic time scale, $\tau_c$.
Accordingly, their stress-relaxation response has two distinct regimes.  
On short time scales ($f\gg 1/\tau_c$), they behave like elastic solids with a storage modulus that is nearly frequency-independent.
On long-time scales ($f\lesssim 1/\tau_c$), they behave like simple fluids with storage, $G'$, and loss, $G''$, moduli that scale as $G'\sim f^2$ and $G''\sim f$, respectively \cite{Rubinstein, Grindy2015,Conrad2019,Katashima2022,Yesilyurt2017, Rosales2020}. 
Materials that depart significantly from this simple behavior can exhibit what is known as ``power-law rheology'': their $G'(f)$ and $G''(f)$ scale with exponents that differ significantly from 2 and 1, respectively  \cite{Rubinstein,Sollich97,Yucht2013,Grindy2015,Dennison2016,FernandezNS2018,SongNonMaxwell}.
Power-law rheology is routinely attributed to the presence of multiple modes of relaxation in the material \cite{Chase2010,Grindy2015,Konuray2023}, but identifying those modes and understanding how they dictate specific power laws remains an open question.

Power-law rheology has been observed in a wide variety of simulated and real materials, including transient hydrogels, and with an equally wide variety of scaling exponents  \cite{Muthu1985,Chambon87,Grindy2015,Kalow2020,Konuray2023,Chase2010,Ward_Weins_Pollak_Weitz_2008,Eberle2012,Lalitha2018}. 
Recently, a $G'\sim G''\sim f^{1/2}$ scaling observed at low frequencies in transiently cross-linked networks of semi-flexible polymers has been explained by a cross-link-governed dynamics model in which the critical frequency is set by the lifetime of the cross-linking bond, $\tau_c\equiv1/f_c$  \cite{Chase2010,Ward_Weins_Pollak_Weitz_2008}.
To explore the origin of power-law exponents in a larger variety of systems, we here use the designability of DNA sequence-based interactions to systematically vary the networked interactions of a transient hydrogel.

\begin{figure}
\centering
\includegraphics[width = 3.375in]{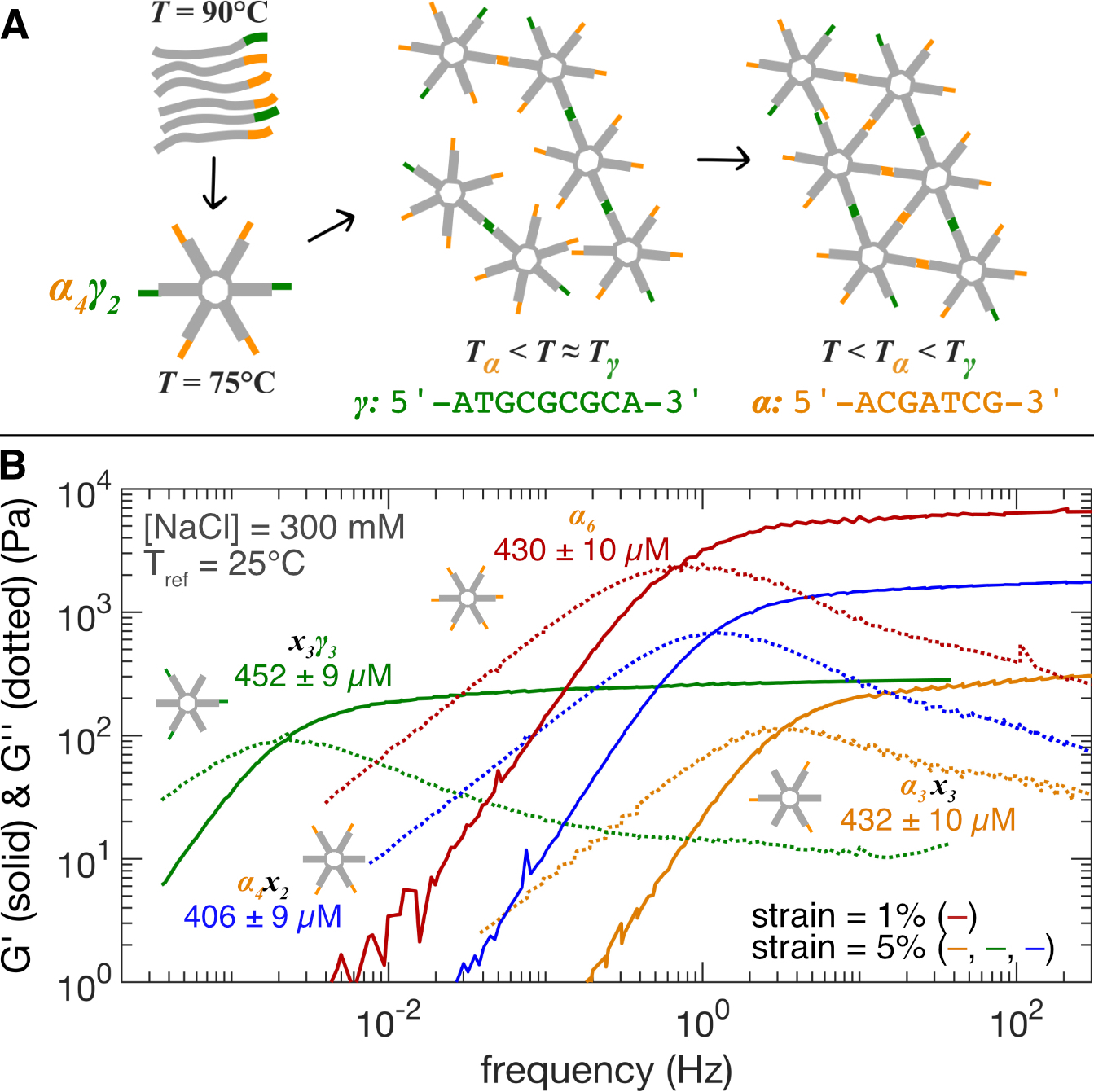}
\caption{\textbf{(A)} Schematic of NS formation and network assembly as temperature, $T$, is lowered.
$T_\alpha$ and $T_\gamma$ denote the predicted melting temperatures of $\alpha$ and $\gamma$ overhangs, respectively. 
Note that co-localization afforded by the $\gamma$ network facilitates $\alpha$ bond formation below $T_\alpha$.
\textbf{(B)} Maxwell-like frequency dependence of the storage ($G'$) and loss ($G''$) moduli of homotypic 6-arm NS hydrogels $\alpha_3 x_3$ (orange), $\alpha_4 x_2$ (blue), $x_3 \gamma_3$ (green), and $\alpha_6$ (red).
}
\label{fig:fig1}
\end{figure}

Our gels are made of DNA nanostars (NS), star-like supramolecules of limited valence that form transient networks via Watson-Crick base-pairing.
To avoid changing the mechanics of the network junctions, we focused on NS that shared the same, six-armed structure, but with different palindromic $5^\prime$ ssDNA overhangs mediating NS-NS binding (Fig. \ref{fig:fig1}A, see Fig. S1 for sequences). 
Each arm had either a weak palindromic sequence ($\alpha$: 5$'$-$\tt{CGATCG}$-3$'$), a strong one ($\gamma$: 5$'$-$\tt{TGCGCGCA}$-3$'$), or none at all ($x$: {\em i.e.}, a blunt end).
Hereafter, we refer to NS with $m$ $\alpha$-overhangs and $n$ $\gamma$-overhangs or $n$ absent overhangs as $\alpha_m \gamma_n$ or $\alpha_m x_n$, respectively.

All NS were self-assembled by mixing their component HPLC-purified ssDNA oligomers in equal concentrations in pure water, dehydrating the mixture, rehydrating it to a final NS concentration between $400~\mu$M and $470~\mu$M in Tris-buffered saline solution (300 mM NaCl, 40 mM Tris, pH 8.3), and, finally, annealing from $90^{\circ}$C to room temperature over $\sim 5$ h.
Once annealed, such concentrated NS solutions form hydrogels that are too viscous to pipette, so a slant-cut pipette tip was used to scoop and place $\sim100~\mu$L on the stationary plate of a strain- and temperature-controlled rheometer (TA Instruments, Ares-G2).

Oscillatory rheology was performed in the linear viscoelastic regime ($\sim1\%$ strain) over a range of frequencies (0.1-10 Hz) and temperatures ($5^{\circ}$C $\leq T\leq50^{\circ}$C).
All $G'$ and $G''$ master curves were constructed using time-temperature superposition (TTS) with a reference temperature of 25$^{\circ}$C, as detailed in \cite{Conrad2019}.
Similarly, oscillatory strain sweep measurements of $G'$ and $G''$ were performed over a range of strains ($1-300\%$) at temperatures and frequencies spanning the non-Newtonian regime of the solution under investigation and
TTS shift factors from the master curves were used to shift the frequencies to the same 25$^{\circ}$C reference temperature.

Hydrogels of 6-arm NS with only one type of NS-NS bond (``homotypic'' NS) were Maxwell-like regardless of the bond strength and even if some arms lacked overhangs (Fig.~\ref{fig:fig1}B).  
Increasing the strength of the NS-NS bond, by changing the overhang sequence from six bases with 66\% GC content ($\alpha$) to eight bases with 75\% GC content ($\gamma$), lowered $f_{c}$ by several orders of magnitude, $f_c(x_3 \gamma_3)\ll f_c(\alpha_3 x_3)$, but had no significant effect on $G'_p$  (Fig.~\ref{fig:fig1}B, orange and green curves).
Increasing NS valence by adding overhangs increased both $\tau_{c}$ and $G'_p$ but the effect on $\tau_{c}$ was less pronounced.
For example, between hydrogels of valence-3 NS ($\alpha_3 x_3$) and those of valence-6 NS ($\alpha_6$) with the same bond strength there was a 5-fold increase in $\tau_{c}$ and a 20-fold increase in $G'_p$ (Fig.~\ref{fig:fig1}B, orange and red curves).
This is similar to our previous results on homotypic NS hydrogels of different arm numbers \cite{Conrad2019}, indicating that blunt-ended arms are rheologically inert. 

\begin{figure}
\centering
\includegraphics[width = 3.375in]{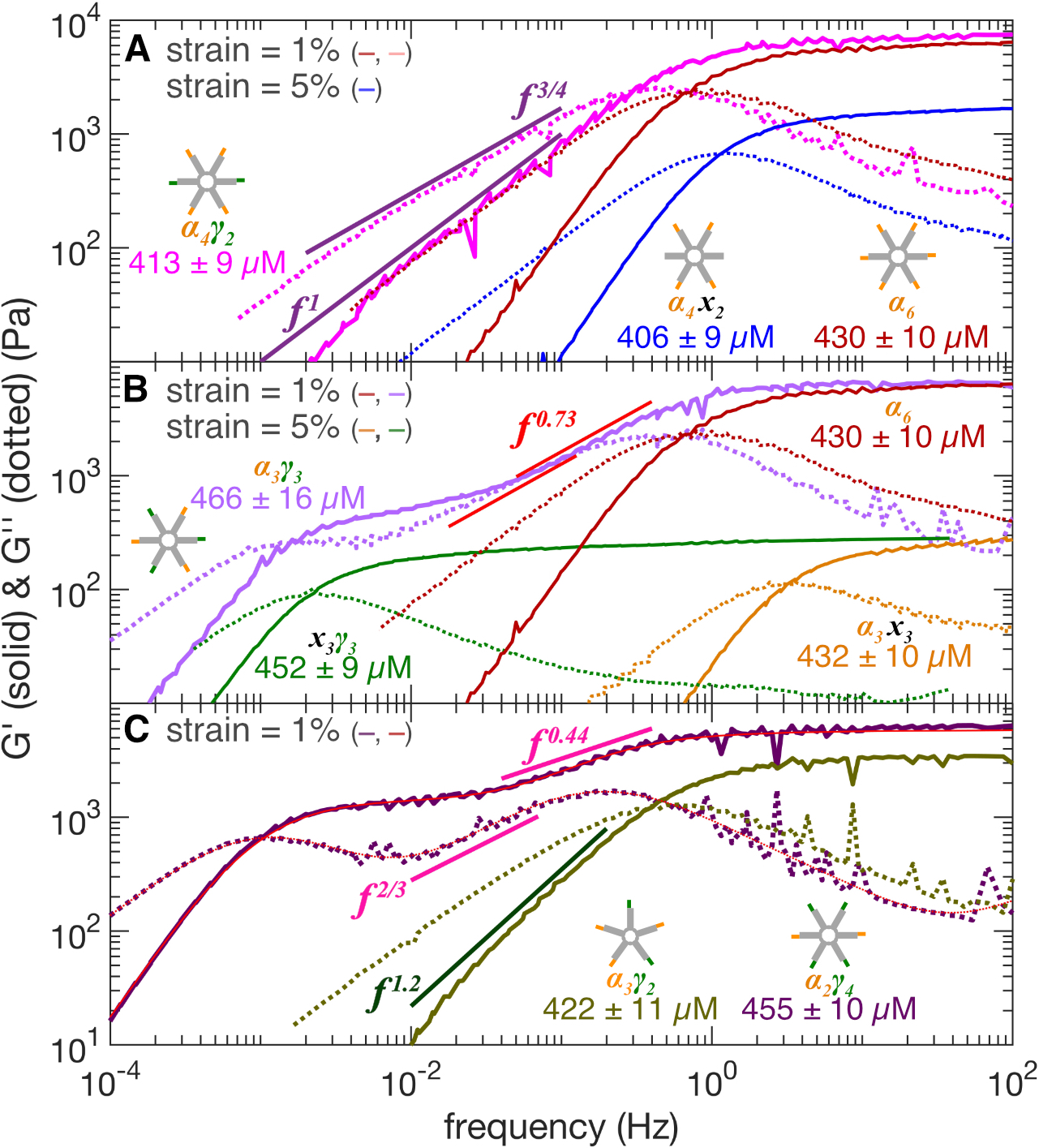}
\caption{Plot of $G'$ and $G''$ versus frequency of hydrogels of \textbf{(A)} $\alpha_4 \gamma_2$ (magenta), $\alpha_4 x_2$ (blue), $\alpha_6$ (red), \textbf{(B)} $x_3 \gamma_3$ (green), $\alpha_3 x_3$ (orange), $\alpha_3 \gamma_3$ (purple), $\alpha_6$ (red), \textbf{(C)} $\alpha_2 \gamma_4$ (dark purple), and $\alpha_3 \gamma_2$ (olive-green) NSs, measured at strains listed in the legends and at $T_\text{ref} =25^{\circ}$C.
Red curves in (C) denote generalized Maxwell fit to $G'$ and $G''$.
}
\label{fig:fig2}
\end{figure}

By contrast, hydrogels of valence-6 NS that made both strong and weak NS-NS bonds (``heterotypic'' NS) exhibited power-law rheology (Fig.~\ref{fig:fig2}).
At all but the very lowest frequencies, their viscoelastic moduli scaled more gradually than those of a Maxwell material and their maximal scaling-exponents diminished with the proportion of weak:strong bonds.
Hydrogels of $\alpha_4 \gamma_2$ NS had $G'\sim f^1$ and $G''\sim f^{3/4}$ below a single, characteristic frequency (at which $dG''/df=0$) that was slightly lower than $f_c(\alpha_6)$ (Fig.~\ref{fig:fig2}A).
Hydrogels of $\alpha_3 \gamma_3$ NS had two characteristic frequencies, $f_1 \lesssim f_c(x_3 \gamma_3) \ll f_2\lesssim f_c(\alpha_6)$ and maximal scaling exponents $\lesssim 3/4$ for both moduli (Fig.~\ref{fig:fig2}B). 
Hydrogels of $\alpha_2 \gamma_4$ NS also had two characteristic frequencies, slightly below the corresponding frequencies of $\alpha_3 \gamma_3$ and maximal scaling exponents $<1/2$ and $<2/3$, for $G'$ and $G''$ respectively (Fig.~\ref{fig:fig2}C). 

One clue to the origin of this power-law rheology is that the storage moduli of all 6-arm heterotypic NS hydrogels plateau to a value similar to that of the $\alpha_6$ NS hydrogel, $G'_p\approx 6500$ Pa. 
This suggests that, as frequency rises, the strong-bond network becomes constrained by an increasing number of weak bonds. 
Another clue is that the moduli of heterotypic NS hydrogels are everywhere larger than those of the homotypic NS hydrogel with a corresponding strong-bond network ({\em e.g.}, $G'_{\alpha_3 \gamma_3} > G'_{x_3 \gamma_3},\,G''_{\alpha_3 \gamma_3} > G''_{x_3 \gamma_3}(f)$).
The moduli are enhanced even at frequencies where the corresponding weakly-bonded homotypic NS hydrogel ({\em e.g.}, $\alpha_3 x_3$) has negligible viscosity. 
This suggests that being inextricably linked by the NS design enables the $\gamma$-bonded network to catalyze $\alpha$-bond formation.

\begin{figure}[t]
\centering
\includegraphics[width = 3.375in]{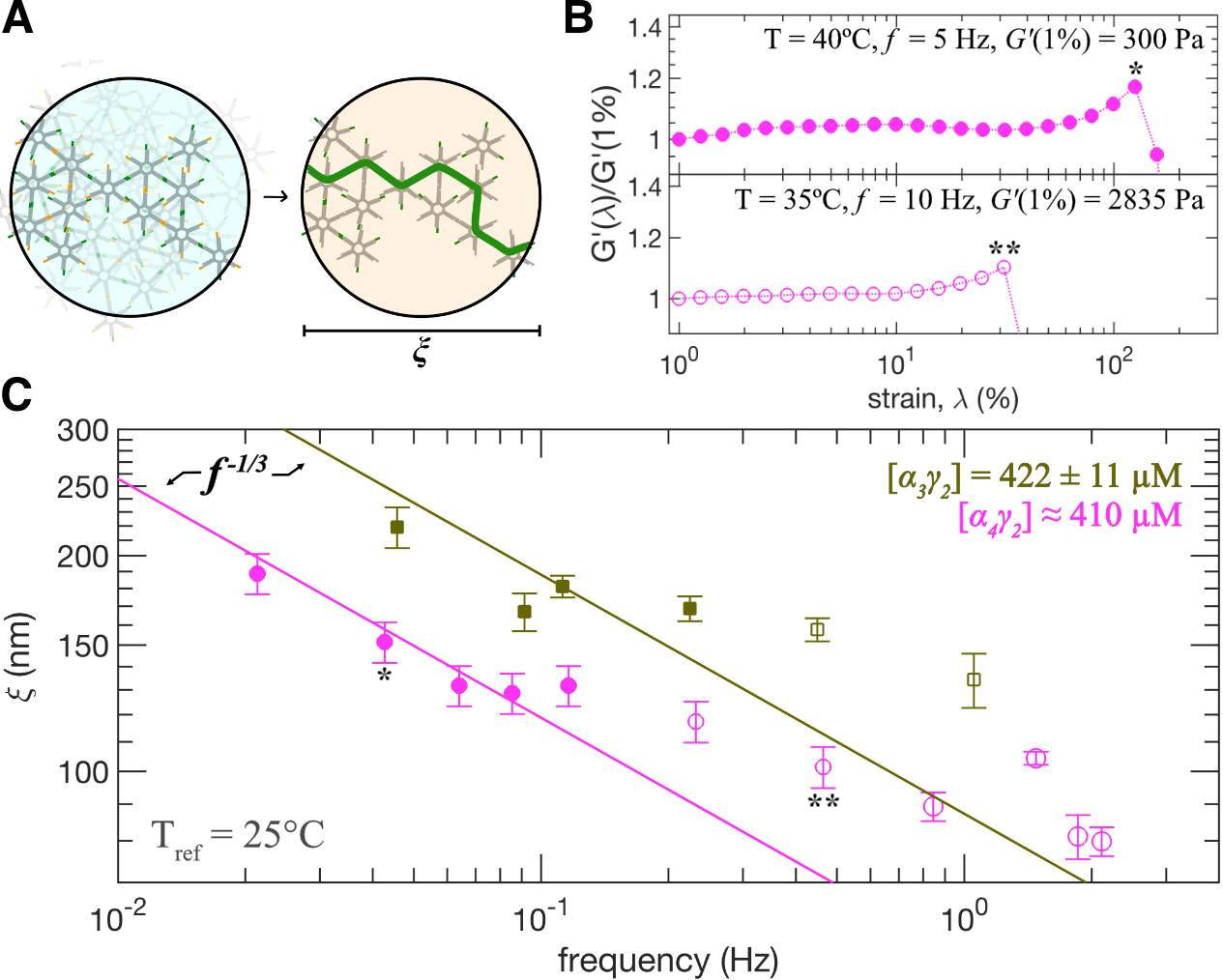}
\caption{(\textbf{A}) Cartoon depiction of the effective medium model showing weak bonds (orange) as viscous solvent through which the dominant stress-bearing chain (bold green path) of the strong-bond network must diffuse to relax.
(\textbf{B}) Examples of strain sweeps used for yield stress estimation, with original temperature, frequency, and storage modulus at $1\%$ strain specified.
(\textbf{C}) Elastic correlation length, $\xi$, estimated from strain sweeps on $\alpha_4\gamma_2$ (pink) and $\alpha_3 \gamma_2$ (olive green) NS hydrogels. 
Closed and open symbols respectively denote when \emph{both} $G'$ and $G''$ do or do not exhibit a power-law scaling with frequency.
Lines are one-parameter fits to $\xi \sim f^{-1/3}$, for comparison to data that lie within the power-law rheology regime (closed symbols). 
Symbols with asterisks correspond to the strain sweeps shown in (B).
The $\alpha_4\gamma_2$ data combines measurements with small concentration differences $\sim10~\mu$M.
}
\label{fig:fig3}
\end{figure}

From these clues, and inspired by the effective medium theory of non-linear rheology in transiently cross-linked metallo-supramolecular networks \cite{Lalitha2018}, we propose to treat the weakly-bonded $\alpha$-network as an effective medium of viscosity, $\eta_\alpha$, and the strongly-bonded $\gamma$-network as a network of units the size of an elastic correlation length, $\xi$, where $\xi > a$, the size of a single NS (Fig. \ref{fig:fig3}A).
We further propose that stress-relaxation in heterotypic NS hydrogels is a diffusive process ({\em i.e.}, the hydrogel relieves stress as $\xi$-sized segments of the strong-bond network diffuse through the effective viscosity created by the weak-bonds). 

To build a diffusive stress-relaxation (DSR) model, we begin by noting that the time required for stress dissipation should scale as $t \sim \xi^2/D(\xi)$, where the diffusion constant $D(\xi)$ depends on the extent of hydrodynamic coupling.  
When hydrodynamic coupling is strong, the so-called ``Zimm'' limit, the $\alpha$-bond medium within a $\xi$-sized region of the strong-bond network cannot drain ({\em i.e.}, re-arrange) as the network diffuses in response to stress. 
The diffusion constant of the stress-bearing chain therefore scales like that of a solid, $\xi$-sized sphere in the effective medium, 
$D_{\xi} \sim k_\mathrm{B}T/\eta_\alpha \xi$, 
where $k_\mathrm{B}$ is the Boltzmann constant and $T$ is the absolute temperature \cite{Rubinstein}.
When the hydrodynamic coupling is so weak that the medium is freely draining, the so-called ``Rouse'' limit, a $\xi$-sized segment drags along only $\gamma$-bonded NS as it diffuses, so 
$D_{\xi} \sim k_\mathrm{B}T/\eta_\alpha N a$, 
where $N \sim (\xi/a)^{d_\text{fr}}$ is the number of $\gamma$-bonded NS in a $\xi$-sized sphere and $d_\text{fr}<3$ 
is the fractal dimension of the $\gamma$-network on timescales that exhibit power-law rheology

Substituting these scaling relations into the diffusion-time equation and solving for $\xi$ yields
\begin{equation}
   \xi \sim 
    \begin{cases}
         \left(\frac{ k_\mathrm{B}T}{\eta_\alpha} t\right)^{1/3} \sim f^{-1/3}&  ,\quad \text{Zimm}\\
        \left(\frac{ k_\mathrm{B}T}{\eta_\alpha a}a^{d_\mathrm{fr}} t\right)^{1/(2+d_\text{fr})}\sim f^{-1/(2+d_\text{fr})}& ,\quad \text{Rouse}
    \end{cases}
\label{eqn:xi(f)meanfield}
\end{equation}  

To derive the scaling relation for $G'(f)$, we recall that the storage modulus of a three-dimensional network goes as $G'\sim K_{\xi}/\xi$, where $K_\xi$ is the spring constant of the dominant stress-bearing chain within $\xi$ \cite{Mellema,Shih,Dages2022}.
Because each stress-bearing chain consists of $N_{\xi} \sim \xi^{d_\text{min}}$ elastic elements in series, where $d_\text{min}$ is the fractal dimension of that shortest elastic chain within $\xi$, the spring constant scales as $K_{\xi} \sim 1/N_{\xi} \sim \xi^{-d_\text{min}}$ \cite{Daoud}. 
Combining these relationships with Eq.~\ref{eqn:xi(f)meanfield} yields
\begin{equation}
    G^\prime \sim \xi^{-(1+d_\text{min})} \sim 
    \begin{cases}
        f^{(1+d_\text{min})/3}& ,\quad \text{Zimm}\\
        f^{(1+d_\text{min})/(2+d_\text{fr})}& ,\quad \text{Rouse}
    \end{cases}
    \label{eqn:G'(f)}
\end{equation}

For two of the three heterotypic NS hydrogels described above, based on what is known of the fractal dimension of stress-bearing chains in their strong-bond networks, the observed power-laws are consistent with Eq.~\ref{eqn:G'(f)} in the Zimm limit.  
In particular, $G'_{\alpha_4 \gamma_2} \sim f^1$ can only be understood in the Zimm limit because, as some nanostars within $\xi$ are not stress-bearing,  $d_\mathrm{min}$ must be less than or equal to $d_\mathrm{fr}$, making it impossible for the Rouse limit to support a scaling exponent of $G'$ that is $>0.8$.
Furthermore, in the Zimm limit, $G'_{\alpha_4 \gamma_2} \sim f^1$ implies $d_\text{min} = 2$, which is the fractal dimension of the freely-jointed chain structure expected of the ``valence-2'' $\gamma$-network in the $\alpha_4 \gamma_2$ NS hydrogel (Fig.~\ref{fig:fig2}A).
For a valence-3 $\gamma$-network in an $\alpha_3 \gamma_3$ NS hydrogel, $d_\text{min} = 1.2$ was estimated from rheology of a three-arm homotypic NS hydrogel \cite{Conrad2019}.
This implies  $G'_{\alpha_3 \gamma_3} \sim f^{0.73}$ in the Zimm limit, in good agreement with the scaling observed between the characteristic frequencies (Fig.~\ref{fig:fig2}B).

The power-law rheology of the $\alpha_2 \gamma_4$ hydrogel, by contrast, cannot be explained in the Zimm limit. 
Its maximal $G'$-scaling 
has an exponent $0.4 \leq q \lesssim0.5$ that, in the Zimm limit of Eq.~\ref{eqn:G'(f)}, implies an unphysical $d_\text{min} \lesssim 0.5$. 
We considered whether $\alpha_2 \gamma_4$ might be better characterized as a simple, bi-modal material with two very different bond lifetimes.  
The generalized Maxwell fit to $G'_{\alpha_2 \gamma_4}(f)$ and $G''_{\alpha_2 \gamma_4}(f)$ does yield a bi-modal relaxation spectrum, with one mode $\approx 400$x slower than the other, but between the modes, it is not consistent with simple bi-modality (Fig.~S3).

We therefore suggest that, despite being intrinsically coupled to the $\gamma_4$ network, the $\alpha_2$ network is too tenuous to support strong hydrodynamic coupling.
In the Rouse limit of the DSR model, the observed $G'$-scaling requires $d_\text{fr} > 2$ and $1 \leq d_\text{min} < 1.5$. 
Prior rheology of a four-arm homotypic NS hydrogel \cite{Conrad2019} doesn't provide an estimate of $d_\text{min}$, but it does suggest a percolated network structure and, therefore, $d_\text{fr} \approx 2.5$. 
This value of $d_\text{fr}$ is not only consistent with the Rouse-limit, it further constrains $1 \leq d_\text{min} \leq 1.25$ to be similar to, or even less than, that of a valence-3 network.
It may be that the tendency of 4-armed DNA junctions to maximize base-stacking by adopting a planar, X-like geometry \cite{Spinozzi20} prevents stress-bearing chains from filling space as efficiently as in a valence-3 network.

To test the DSR model further, we performed rheology on hydrogels of a 5-armed NS: $\alpha_3 \gamma_2$ (Fig. \ref{fig:fig3}C). 
Intriguingly, below a single characteristic frequency $f_c(\alpha_3 \gamma_2) \approx f_c(\alpha_4 \gamma_2)$, the scaling of $G'_{\alpha_3 \gamma_2} \sim f^{1.2}$ was steeper than that observed in $\alpha_4 \gamma_2$.
Given that such steep scaling is only possible in the Zimm limit, Eq.~\ref{eqn:G'(f)} implies $d_\text{min} = 2.6$.
We speculate that in the context of a 5-armed NS, the valence-2 $\gamma$-network is likely to kink at the junction and therefore fills space more efficiently than a freely jointed chain.
Future work might test this hypothesis by studying $\alpha_3 x_1 \gamma_2$ hydrogels.
When the $\gamma$-arms of such a NS are diametrically opposed, the power-law rheology should be indistinguishable from that of $\alpha_4 \gamma_2$ above (in which the $\gamma$-arms were so-arranged), but when the $\gamma$-arms share a common strand, the $G'(f)$ scaling should be steeper than that of $\alpha_3 \gamma_2$, whose more flexible junction imposes less intrinsic curvature on the strong-bond network.

The DSR model also predicts the frequency-dependence of $\xi$, which can be deduced from strain-sweep measurements of the yield stress (Fig. \ref{fig:fig3}B).
Specifically, by measuring the yield strain, $\lambda_y$, and yield storage modulus, $G'_y$, it is possible to estimate the yield stress, $\sigma_y = G'_y\lambda_y$. 
Because $\sigma_y$ scales as the force required to break a bond in the strong-bond network, $F_c$ ($\approx10$~pN based on the $\gamma$-overhang sequence \cite{HO20093158}), divided by the characteristic area between such bonds, $\xi^2$, 
we expect
\begin{equation}
 \sigma_y \equiv G^\prime_y\lambda_y\sim F_c / \xi^2 \implies \xi \sim \sqrt{\frac{F_c}{G'_y\lambda_y }}\,,
 \label{eqn:xi(f)strainsweep}
\end{equation}
In $\alpha_3 \gamma_3$ and $\alpha_2 \gamma_4$ NS hydrogels, the frequency range of $G'$-scaling was too narrow and the precision of our $\xi$ measurement was too low ($\sim20\%$) to meaningfully constrain the scaling of $\xi(f)$. 
In $\alpha_4 \gamma_2$ and $\alpha_3 \gamma_2$ NS hydrogels, however, the data was reasonably consistent with $\xi \sim f^{-1/3}$ over the near-decade in frequency for which their steep $G'$-scaling places them firmly in the Zimm limit of the DSR model  (Eq.~\ref{eqn:xi(f)meanfield}).
Because it relies on a data set that is wholly orthogonal to  linear elastic rheology (Fig.~\ref{fig:fig2}); this result provides strong support for the validity of the DSR picture.

An even stronger test of the model would be direct measurement of $d_{\text{frac}}$ and $d_{\text{min}}$ in heterotypic NS solutions.
It may be feasible to measure $d_{\text{frac}}$ via scattering, but acquiring sufficient signal with integration times relevant to the power-law rheology regime will be challenging. 
It is likely not feasible to measure $d_{\text{min}}$ directly because there is no obvious way to distinguish between stress-bearing and non-stress-bearing NS.
We looked, instead, to simulations of network dynamics, from which it is possible to measure both $G'(f)$ and $d_{\text{min}}$.  

Several simulations of two-dimensional networks report $G'(f)$ and $d_{\text{min}}$ \cite{Chase2010,Yucht2013, Dennison2016}.
The permanently-bonded, near-isostatic network in these simulations plays the role of the strong-bond network and the mean-field viscosity plays the role of the transient, weak-bond network in the heterotypic NS hydrogels that inspired the DSR model. 
While $\xi(f)$ scales the same in $2d$ as in $3d$, the scaling of $G'$ with $\xi$ is different.
For a two-dimensional network, instead of depending inversely on $\xi$, the storage modulus is directly proportional to the stress-bearing chain's spring constant, $G' \sim K_\xi$. 
Therefore, in $2d$ the DSR model predicts:
\begin{equation}
   G'_{2d}  \sim \xi^{-d_\text{min}} \sim 
    \begin{cases}
       f^{d_\text{min}/3}& ,\quad  \text{Zimm}\\
       f^{d_\text{min}/(2+d_\text{fr})}& ,\quad \text{Rouse}
    \end{cases}
\label{eqn:G'2d}
\end{equation}  
Simulations of a $2d$ triangular lattice at the isostatic threshold \cite{Thorpe1996}, found $d_{\text{min}} = 1.80 \pm 0.03$ and $d_{\text{fr}} = 1.86 \pm 0.02$.
Plugging these values into Eqns.~\ref{eqn:G'2d}, yields $G^\prime \sim f^{0.60 \pm 0.01}$ and $G' \sim f^{0.47 \pm 0.02}$ in the Zimm- and Rouse-limits, respectively.
These values agree with others' simulations on the same lattice \cite{Yucht2013, Dennison2016}, which measured $G' \sim f^{0.61}$ in the Zimm limit and $G^\prime\sim f^{0.41}$ in the Rouse limit. 

In summary, we leveraged the designability of DNA nanostars to study power-law rheology in transient hydrogels that contain two different bonds, one weak and one strong, and varied the weak:strong bond ratio.
We found that introducing a second bond timescale does not simply add a second macroscopic relaxation mode, but rather creates emergent dynamics with a broad spectrum of relaxation modes.
We show that an effective medium model of diffusive stress-relaxation (DSR) is consistent with the data and validated by agreement with simulations of ($2d$) near-isostatic networks that are themselves consistent with theoretical calculations \cite{Yucht2013,Dennison2016}.
We therefore posit that the power-law exponents scaling $\xi$ and $G'$ with frequency are related to the fractal dimensions of the strong-bond network and of its stress-bearing chains.
Overall, this work demonstrates the use of DNA nanotechnology to decipher, and potentially rationally design, power-law rheology.

\begin{acknowledgments}
We thank Chase Broedersz and Fred MacKintosh for helpful discussions.
This project was supported by NSF Award No. CMMI 1935400 and made use of shared facilities of the UCSB MRSEC (NSF DMR 2308708).
\end{acknowledgments}

\bibliography{DTI.bib}

\end{document}